\documentclass[osajnl,twocolumn,showpacs,superscriptaddress,10pt]{revtex4-1}
\usepackage{amsmath,amssymb,graphicx}

\begin{document}

\title{Nonreciprocal transmission and fast-slow light effects \\
in a cavity optomechanical system}
\author{Jun-Hao Liu}
\author{Ya-Fei Yu}
\email{yuyafei@m.scnu.edu.cn}
\author{Zhi-Ming Zhang}
\email{zhangzhiming@m.scnu.edu.cn}
\address{Guangdong Provincial Key Laboratory of Nanophotonic Functional Materials and Devices 
	(School of Information and Optoelectronic Science and Engineering),
	and Guangdong Provincial Key Laboratory of Quantum Engineering and Quantum Materials,
	South China Normal University, Guangzhou 510006, China} 
\begin{abstract}
We study the nonreciprocal transmission and the fast-slow light effects in a
cavity optomechanical system, in which the cavity supports a clockwise and a
counter-clockwise circulating optical modes, both the two modes are driven
simultaneously by a strong pump field and a weak signal field. We find that
when the intrinsic photon loss of the cavity is equal to the external coupling
loss of the cavity, the system reveals a nonreciprocal transmission of the
signal fields. However, when the intrinsic photon loss is much less than the external
coupling loss, the nonreciprocity about the transmission properties almost
disappears, and the nonreciprocity is shown in the group delay properties of
the signal fields, and the system exhibits a nonreciprocal fast-slow light
propagation phenomenon.
\end{abstract}

\maketitle



\section{Introduction}

In recent years, optical nonreciprocity has got a lot of attentions for its
important applications in photonic network, signal processing, and one-way
optical communication protocols. In the common nonreciprocal devices, such
as, isolator, circulator, nonreciprocal phase shifter, the transmission of
the information is not symmetric. At present, the researches about the
optical nonreciprocity mainly focused on two aspects: one is the
transmission properties of the signal fields, another is the photonic
statistical properties of the signal fields.

For the first aspect, scientists have demonstrated that many physical
effects and physical systems, such as Faraday rotation effect in the
magneto-optical crystals \cite{1,2,3}, optical nonlinearity \cite{4},
spatial-symmetry-breaking structures \cite{5,6}, optoacoustic effects \cite%
{7,8}, the parity-time-symmetric structures \cite{9,10,11,12}, can be used
to realize the optical nonreciprocal transmission. Efforts have also been
made to study the nonreciprocal transmission in cavity optomechanical
systems \cite{13,14,15,16,17}. Manipatruni et al. demonstrated that the
optical nonreciprocal transmission was based on the momentum difference
between the forward and backward-moving light beams in a Fabry-Perot cavity
with one moveable mirror \cite{18}. Hafezi et al. proposed a scheme to
achieve the nonreciprocal transmission in a microring resonator by using an
unidirectional optical pump \cite{19}. Metelmann and Clerk discussed a
general method for nonreciprocal photon transmission and amplification via
reservoir engineering \cite{20}. Peterson et al. demonstrated an efficient
frequency-converting microwave isolator based on the optomechanical
interactions \cite{21}. Mirza et al. studied the optical nonreciprocity and
slow light propagation in coupled spinning optomechanical resonators \cite%
{22}.

For the second aspect, the researches on the photonic statistic properties
of the transmitted fields in nonreciprocal devices are fewer. At present,
the relevant theoretical works include the nonreciprocal photon blockade 
\cite{23}, the authors discussed how to create and manipulate nonclassical
light via photon blockade in rotating nonlinear devices. They found that the
light with sub-Poissonian or super-Poissonian photon-number statistics can
emerge when driving the resonator from its left or right side. Subsequently,
Xu et al. proposed a scheme to manipulate the statistic properties of the
photons transport nonreciprocally via quadratic optomechanical coupling \cite%
{24}.

\begin{figure*}[tbp]
	\centering\includegraphics[width=11cm,height=6.24cm]{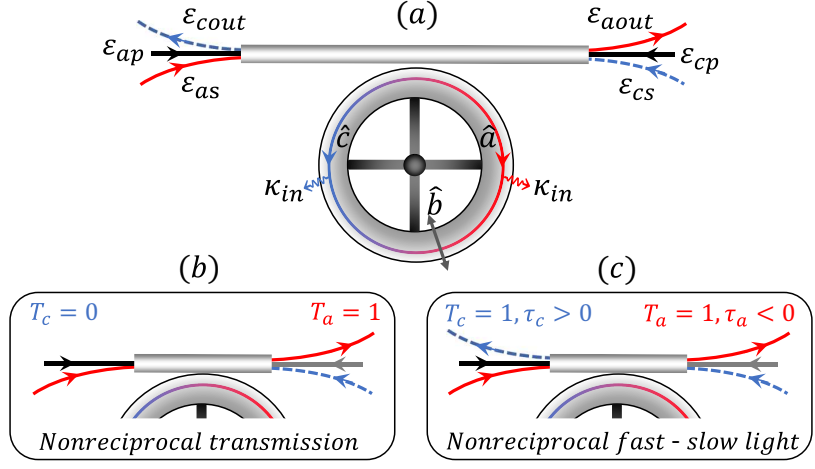}
	\caption{(a) Schematic diagram of our proposed model. An optomechanical
		microtoroid cavity supports a clockwise circulating mode ($\hat{a}$) and a
		counter-clockwise circulating mode ($\hat{c}$), both the two cavity modes 
		couple with the mechanical mode ($\hat{b}$) via the radiation pressure. The pump
		fields ($\protect\varepsilon _{ap}$, $\protect\varepsilon _{cp}$) and signal
		fields ($\protect\varepsilon _{as}$, $\protect\varepsilon _{cs}$) couple
		with the cavity modes by an optical fiber. (b) The nonreciprocal transmission:
		the right-moving signal field is completely transmitted ($T_{a}=1$), while
		the left-moving signal field is blocking-up ($T_{c}=0$). (c) The nonreciprocal
		fast-slow light: both the right-moving field and the left-moving signal field are 
		transmitted ($T_{a}=T_{c}=1$). However, the group delay of the right-moving signal
		field is negative ($\protect\tau_{a}<0$), that corresponds to the fast
		light. The group delay of the left-moving signal field is positive ($\protect%
		\tau_{c}>0$), that corresponds to the slow light. }
\end{figure*}

In this paper, we study the nonreciprocal transmission and the fast-slow light
effects in a cavity optomechanical system, as shown in Fig. 1. We show that
when the intrinsic photon loss of the cavity equals the external coupling
loss of the cavity, we can achieve the nonreciprocal transmission of the
signal fields with the red-sideband pumping or the blue-sideband pumping. We
also show that when the intrinsic photon loss is much less than the
external coupling loss, the nonreciprocity of the system about the optical
transmission properties almost disappears, now the system exhibits a
nonreciprocal fast-slow light propagation phenomenon, i.e., the group
velocity of the right-moving signal field will be speed up (fast light),
while the group velocity of the left-moving signal field will be slowed down
(slow light), or vice versa.

\section{Model and Hamiltonian}

Our system model is shown in Fig. 1(a). We consider an optomechanical
microtoroid cavity, which supports a clockwise circulating mode ($\hat{a}$)
and a counter-clockwise circulating mode ($\hat{c}$), both the two cavity modes
 couple with the mechanical mode ($\hat{b}$) via the radiation pressure. The cavity mode $\hat{a}$ ($\hat{c}$) is driven simultaneously by a
strong pump field $\varepsilon _{ap}$ ($\varepsilon _{cp} $) and a weak
signal field $\varepsilon _{as}$ ($\varepsilon _{cs}$). The total
Hamiltonian of the system can be expressed as 
\begin{equation}
H_{total}=H_{om}+H_{aps}+H_{cps}+H_{ac},
\end{equation}%
where $H_{om}=\hbar \omega _{0}\hat{a}^{\dag }\hat{a}+\hbar \omega _{0}\hat{c%
}^{\dag }\hat{c}+\hbar \omega _{m}\hat{b}^{\dag }\hat{b}+\hbar g(\hat{a}%
^{\dag }\hat{a}+\hat{c}^{\dag }\hat{c})(\hat{b}^{\dag }+\hat{b})$ is the
Hamiltonian of the cavity optomechanical system. $\hat{a}(\hat{c})$ and $%
\hat{b}$ are the annihilation operators of the clockwise (counter clockwise)
circulating cavity mode and the mechanical mode with frequency $\omega _{0}$
and $\omega _{m}$, respectively. $g$ is the optomechanical coupling strength
between the cavity modes and the mechanical mode. $H_{aps}=i\hbar
\varepsilon _{ap}(\hat{a}^{\dag }e^{-i\omega _{ap}t}-H.c.)+i\hbar
\varepsilon _{as}(\hat{a}^{\dag }e^{-i\omega _{as}t}-H.c.)$ describes the
interactions of the cavity mode $\hat{a}$ with the pump field of amplitude $%
\varepsilon _{ap}=\sqrt{2\kappa P_{ap}/\hbar \omega _{ap}}$ and the signal
field of amplitude $\varepsilon _{as}=\sqrt{2\kappa P_{as}/\hbar \omega _{as}%
}$, respectively, in which $\kappa $ is the coupling decay rate of the
cavity, and $P_{ap}$ ($P_{as}$) is the laser power. Similarly, $%
H_{cps}=i\hbar \varepsilon _{cp}(\hat{c}^{\dag }e^{-i\omega
	_{cp}t}-H.c.)+i\hbar \varepsilon _{cs}(\hat{c}^{\dag }e^{-i\omega
	_{cs}t}-H.c.)$ describes the interaction Hamiltonian of cavity mode $\hat{c}$
with the pump field of amplitude $\varepsilon _{cp}=\sqrt{2\kappa
	P_{cp}/\hbar \omega _{cp}}$ and the signal field of amplitude $\varepsilon
_{cs}=\sqrt{2\kappa P_{cs}/\hbar \omega _{cs}}$, respectively. The last term 
$H_{ac}=\hbar J(\hat{a}^{\dag }\hat{c}+\hat{c}^{\dag }\hat{a})$ represents
the interaction between the two cavity modes with the strength $J$.

For simplicity, we assume that the two pump fields have the same frequency,
i.e., $\omega _{ap}=\omega _{cp}=\omega _{p}$. In the rotation frame with $%
H_{r}$ $=$ $\omega _{p}(\hat{a}^{\dag }\hat{a}+\hat{c}^{\dag }\hat{c})$, the
system Hamiltonian can be written as 
\begin{eqnarray}
H &=&\hbar \Delta \hat{a}^{\dag }\hat{a}+\hbar \Delta \hat{c}^{\dag }\hat{c}%
+\hbar \omega _{m}\hat{b}^{\dag }\hat{b}+\hbar g(\hat{a}^{\dag }\hat{a}+\hat{%
	c}^{\dag }\hat{c})(\hat{b}^{\dag }+\hat{b})  \notag \\
&&+i\hbar \varepsilon _{ap}(\hat{a}^{\dag }-H.c.)+i\hbar \varepsilon _{as}(%
\hat{a}^{\dag }e^{-i\delta _{as}t}-H.c.)  \notag \\
&&+i\hbar \varepsilon _{cp}(\hat{c}^{\dag }-H.c.)+i\hbar \varepsilon _{cs}(%
\hat{c}^{\dag }e^{-i\delta _{cs}t}-H.c.)  \notag \\
&&+\hbar J(\hat{a}^{\dag }\hat{c}+\hat{c}^{\dag }\hat{a}),
\end{eqnarray}%
where $\Delta =\omega _{c}-\omega _{p}$ is the frequency detuning between
the cavity field ($\hat{a},\hat{c}$) and the pump field ($\varepsilon _{ap}$%
, $\varepsilon _{cp}$), and $\delta _{as}=\omega _{as}-\omega _{p}$ ($\delta
_{cs}=\omega _{cs}-\omega _{p}$) is the frequency detuning between the
signal field $\varepsilon _{as}$ ($\varepsilon _{cs}$) and the pump field $%
\varepsilon _{ap}$ ($\varepsilon _{cp}$). The system dynamics is fully
described by the set of quantum Heisenberg-Langevin equations 
\begin{eqnarray}
\frac{d\hat{a}}{dt} &=&-(i\Delta +\kappa _{t})\hat{a}-ig\hat{a}(\hat{b}%
^{\dag }+\hat{b})-iJ\hat{c}+\varepsilon _{ap}  \notag \\
&&+\varepsilon _{as}e^{-i\delta _{as}t}+\sqrt{2\kappa }\hat{a}_{in},  \notag
\\
\frac{d\hat{c}}{dt} &=&-(i\Delta +\kappa _{t})\hat{c}-ig\hat{c}(\hat{b}%
^{\dag }+\hat{b})-iJ\hat{a}+\varepsilon _{cp}  \notag \\
&&+\varepsilon _{cs}e^{-i\delta _{cs}t}+\sqrt{2\kappa }\hat{c}_{in},  \notag
\\
\frac{d\hat{b}}{dt} &=&-(i\omega _{m}+\gamma )\hat{b}-ig(\hat{a}^{\dag }\hat{%
	a}+\hat{c}^{\dag }\hat{c})+\sqrt{2\gamma }\hat{b}_{in},
\end{eqnarray}%
where the cavity has the damping rate $\kappa _{t}=\kappa _{in}+\kappa $,
which are assumed to be due to the intrinsic photon loss and external
coupling loss, respectively, and the mechanical mode has the damping rate $%
\gamma $. $\hat{a}_{in}$ ($\hat{c}_{in}$), and $\hat{b}_{in}$ are the $\delta $%
-correlated operators of the input noises for the cavity mode $\hat{a}$ ($%
\hat{c}$) and the mechanical mode $\hat{b}$, respectively. These noise operators
satisfy $\left\langle \hat{a}_{in}\right\rangle =\left\langle \hat{c}%
_{in}\right\rangle =\langle \hat{b}_{in}\rangle =0$.

In this model, we are interested in the mean response of the system. Thus,
in the following, we turn to calculate the evolutions of the expectation
values of $\hat{a}$, $\hat{c}$, $\hat{b}$, and we denote $\left\langle 
\hat{a}\right\rangle \equiv A$, $\left\langle \hat{c}\right\rangle \equiv C$%
, $\langle \hat{b}\rangle \equiv B$, $\left\langle \hat{a}^{\dag
}\right\rangle \equiv A^{*}$, $\left\langle \hat{c}^{\dag }\right\rangle
\equiv C^{*}$, $\langle \hat{b}^{\dag }\rangle \equiv B^{*}$. By using the
mean-field assumption $\langle \hat{a}\hat{b}\hat{c}\rangle =\left\langle 
\hat{a}\right\rangle \langle \hat{b}\rangle \left\langle \hat{c}%
\right\rangle $, we can write the equations for the mean values as 
\begin{eqnarray}
\frac{dA}{dt} &=&-(i\Delta +\kappa _{t})A-igA(B+B^{*})-iJC  \notag \\
&&+\varepsilon _{ap}+\varepsilon _{as}e^{-i\delta _{as}t},  \notag \\
\frac{dC}{dt} &=&-(i\Delta +\kappa _{t})C-igC(B+B^{*})-iJA  \notag \\
&&+\varepsilon _{cp}+\varepsilon _{cs}e^{-i\delta _{cs}t},  \notag \\
\frac{dB}{dt} &=&-(i\omega _{m}+\gamma )B-ig(\left\vert A\right\vert
^{2}+\left\vert C\right\vert ^{2}).
\end{eqnarray}

Equations (4) can be solved by using the perturbation method in the limit of
the strong pump fields, while taking the signal fields to be weak. Using the
linearization approximation, we make the following ansatz \cite{25} 
\begin{eqnarray}
X &=&X_{0}+X_{a+}e^{-i\delta _{as}t}+X_{a-}e^{i\delta _{as}t}  \notag \\
&&+X_{c+}e^{-i\delta _{cs}t}+X_{c-}e^{i\delta _{cs}t},
\end{eqnarray}%
where $X$ can be any one of the quantities $A$, $B$, $C$, or their complex
conjugates $A^{*}$, $C^{*}$, $B^{*}$. $X_{0}$ represents the steady-state
mean value of the corresponding system mode, and $X_{a+}$, $X_{a-}$, $X_{c+}$%
, $X_{c-}$ are the additional fluctuations. By substituting Eq. (5) into
Eqs. (4), and keeping only the first-order in the small quantities and
neglecting the nonlinear terms like $A_{a+}C_{a+}$, $A_{a+}B_{c-}$, $%
B_{c-}C_{a+}$, $\cdots $, we can obtain the steady-state mean value
equations, and the fluctuation equations for the cavity mode components $%
A_{a+}$ and $C_{c+}$ (see the appendix). By solving these equations, we find
that $A_{a+}=\eta (\delta _{as})\varepsilon _{as}$, $C_{c+}=\xi (\delta
_{cs})\varepsilon _{cs}$, the concrete forms of the coefficients $\eta
(\delta _{as})$ and $\xi (\delta _{cs})$ are tediously long, and we will not
write them out here.

The relation among the input, internal, and output fields is given as \cite%
{26} $X^{out} $ $= $ $X^{in}-2\kappa X$. By using the ansatz again, we write
the output field $X^{out}$ as $X_{0}^{out}$ $+$ $X_{a+}^{out}e^{-i\delta
	_{as}t}$ $+$ $X_{a-}^{out}e^{i\delta _{as}t}$ $+$ $X_{c+}^{out}e^{-i\delta
	_{cs}t}$ $+$ $X_{c-}^{out}e^{i\delta _{cs}t}$. Then we can obtain the output
field components $A_{a+}^{out} $ $= $ $\varepsilon _{as}-2\kappa A_{a+}$ and 
$C_{c+}^{out} $ $= $ $\varepsilon _{cs}-2\kappa C_{c+}$. The
transmissivities can be written as $t_{a}(\delta
_{as})=A_{a+}^{out}/\varepsilon _{as}$, $t_{c}(\delta
_{cs})=C_{c+}^{out}/\varepsilon _{cs}$. The nonreciprocal transmission is
then described by the normalized transmissivities (transmission spectra) 
\begin{eqnarray}
T_{a} &=&\left\vert t_{a}(\delta _{as})\right\vert ^{2}=\left\vert 1-2\kappa
\eta (\delta _{as})\right\vert ^{2},  \notag \\
T_{c} &=&\left\vert t_{c}(\delta _{cs})\right\vert ^{2}=\left\vert 1-2\kappa
\xi (\delta _{cs})\right\vert ^{2}.
\end{eqnarray}

What's more, in the resonant region of the transmission spectra, the output
signal fields have the phase dispersions $\phi _{a}(\omega _{as})$ $=$ $\arg
[T_{a}(\omega _{as})]$ and $\phi _{c}(\omega _{cs})$ $=$ $\arg [T_{c}(\omega
_{cs})]$, which can cause the group delay \cite{27} 
\begin{equation}
\tau _{a}=\frac{d\phi _{a}(\omega _{as})}{d\omega _{as}},\tau _{c}=\frac{%
	d\phi _{c}(\omega _{cs})}{d\omega _{cs}}.
\end{equation}%
The group delay $\tau _{a}$ $(\tau _{c})$ $>$ $0$ corresponds to the slow
light propagation of the signal field, and the group delay $\tau _{a}$ $%
(\tau _{c})$ $<0$ corresponds to the fast light propagation of the signal
field. In the following, we will discuss the nonreciprocal transmission ($%
T_{a}=1$, $T_{c}=0$ or $T_{a}=0$, $T_{c}=1$) and the nonreciprocal fast-slow
light effects ($\tau _{a}>0$, $\tau _{c}<0$ or $\tau _{a}<0$, $\tau _{c}>0$%
), respectively.

In this paper, the parameters are chosen based on the recently experiment%
\cite{28,29}: $\omega _{m}$ $=$ $2\pi \times 10$ MHz and $\gamma $ $=$ $2\pi
\times 10^{2}$ Hz (quality factor $Q_{m}$ $=$ $10^{5}$), the equivalent mass
of the mechanical resonance $m=5$ ng, and the equivalent cavity length $l=1$
mm. The damping rate of the optical cavity $\kappa $ $=$ $2\pi \times 1$
MHz, the wavelength of the pump laser $\lambda =1064$ nm. The other
parameters are $J$ $=$ $2\pi \times 10^{3}$ Hz, $\kappa _{in}$ $=$ $2\pi
\times 1$ MHz. 
\begin{figure}[b]
	\centering\includegraphics[width=7.5cm,height=8.58cm]{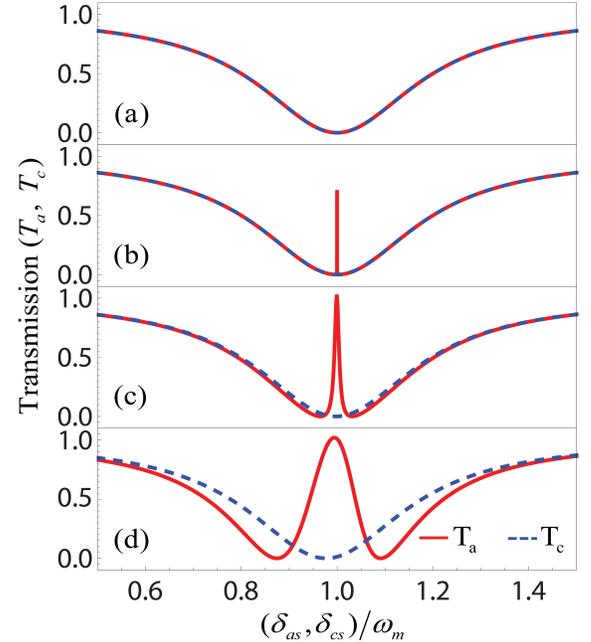}
	\caption{The transmission spectra $T_{a}$ (red solid lines) and $T_{c}$
		(blue dashed lines) as a function of $\protect\delta _{as}/\protect\omega %
		_{m}$ and $\protect\delta _{cs}/\protect\omega _{m}$, respectively. The
		system works near the red sideband ($\Delta =\protect\omega _{m}$). The
		parameters are: (a) $P_{a}$ $=$ $P_{c}$, (b) $P_{a}$ $=$ $10^{2}P_{c}$, (c) $%
		P_{a}$ $=$ $10^{4}P_{c}$, (d) $P_{a}$ $=$ $10^{5}P_{c}$. The other
		parameters are stated in the text.}
\end{figure}
\begin{figure}[tph]
	\centering\includegraphics[width=7.5cm,height=12.34cm]{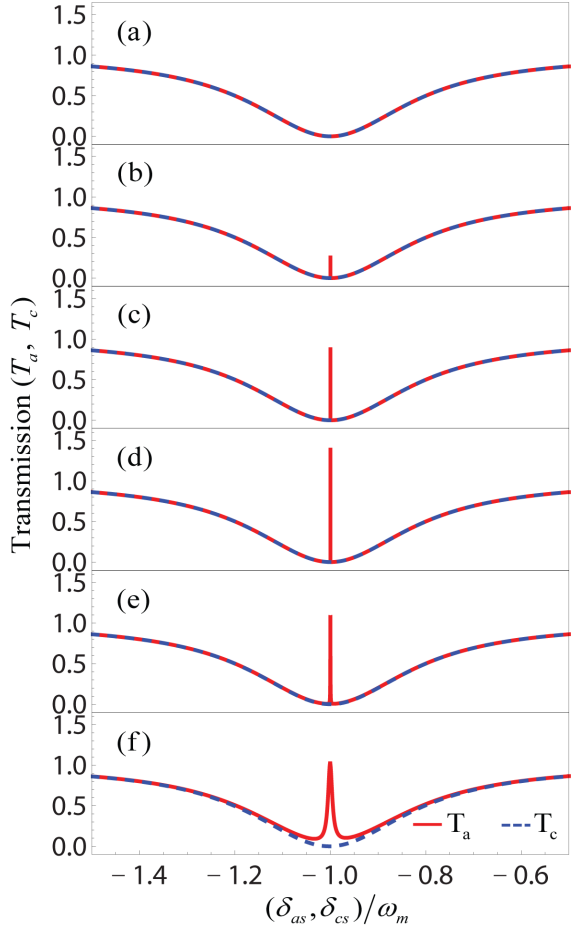}
	\caption{The transmission spectra $T_{a}$ (red solid lines) and $T_{c}$
		(blue dashed lines) as a function of $\protect\delta _{as}/\protect\omega %
		_{m}$ and $\protect\delta _{cs}/\protect\omega _{m}$, respectively. The
		system works near the blue sideband ($\Delta =-\protect\omega _{m}$). The
		parameters are: (a) $P_{a}$ $=$ $P_{c}$, (b) $P_{a}$ $=$ $6P_{c}$, (c) $%
		P_{a} $ $=$ $8.5P_{c}$, (d) $P_{a}$ $=$ $9.5P_{c}$, (e) $P_{a}$ $=$ $5\times
		10^{2} P_{c}$, (f) $P_{a}$ $=$ $10^{4}P_{c}$. The other parameters are
		stated in the text.}
\end{figure}

\section{Nonreciprocal transmission}

In this section, we numerically evaluate the transmission spectra $T_{a}$
and $T_{c}$ to show the possibility of achieving the nonreciprocal
transmission of the signal fields.

\begin{figure}[b]
	\centering\includegraphics[width=7.5cm,height=8.5cm]{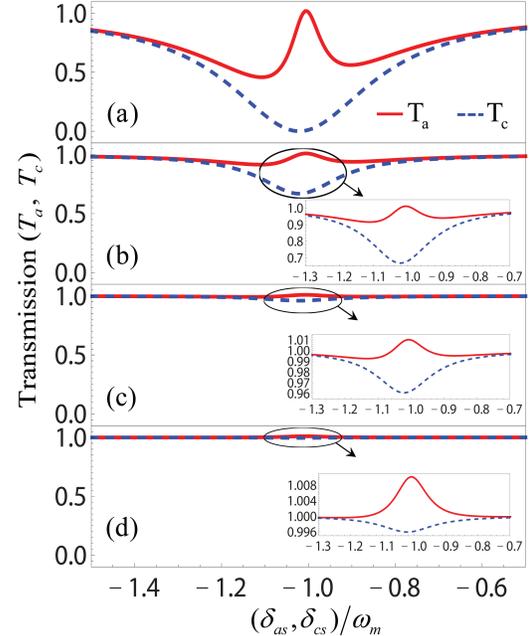}
	\caption{The transmission spectra $T_{a}$ (red solid lines) and $T_{c}$
		(blue dashed lines) as a function of $\protect\delta _{as}/\protect\omega %
		_{m}$ and $\protect\delta _{cs}/\protect\omega _{m}$, respectively, for
		different intrinsic photon loss rate $\protect\kappa _{in}$. The system
		works near the blue sideband ($\Delta =-\protect\omega _{m}$). The
		parameters are: (a) $\protect\kappa _{in}$ $=$ $\protect\kappa$, (b) $%
		\protect\kappa _{in}$ $=$ $10^{-1}\protect\kappa$,, (c) $\protect\kappa %
		_{in} $ $=$ $10^{-2}\protect\kappa$, (d) $\protect\kappa _{in}$ $=$ $10^{-3}%
		\protect\kappa$. The other parameters are stated in the text.}
\end{figure}

\begin{figure}[t]
	\centering\includegraphics[width=7.5cm,height=10.13cm]{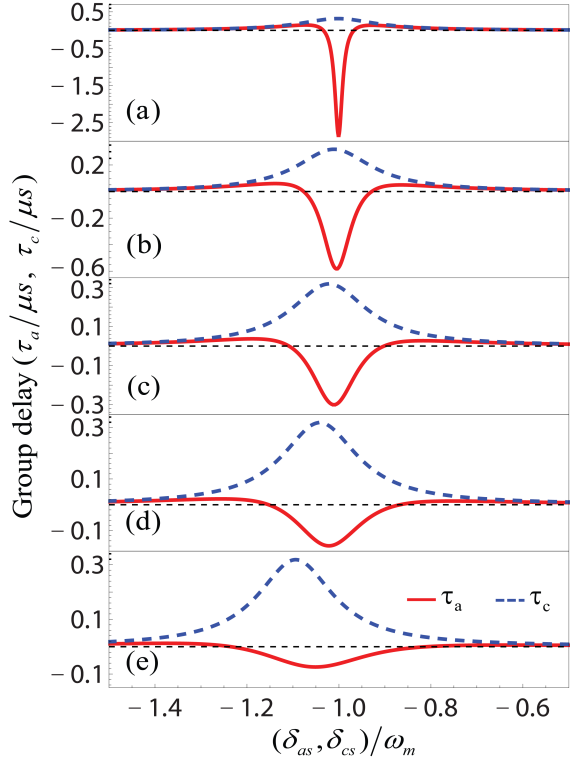}
	\caption{The group delay $\protect\tau_{a}$ (red solid lines) and $\protect%
		\tau_{c}$ (blue dashed lines) as a function of $\protect\delta _{as}/\protect%
		\omega _{m}$ and $\protect\delta _{cs}/\protect\omega _{m}$, respectively.
		The system works near the blue sideband ($\Delta =-\protect\omega _{m}$).
		The parameters are: (a) $P_{a}$ $=$ $1\times 10^{4} P_{c}$, (b) $P_{a}$ $=$ $%
		5\times 10^{4} P_{c}$, (c) $P_{a}$ $=$ $1\times 10^{5} P_{c} $, (d) $P_{a}$ $%
		=$ $2\times 10^{5} P_{c}$, (e) $P_{a}$ $=$ $5\times 10^{5} P_{c}$. The other
		parameters are stated in the text.}
\end{figure}

Firstly, we assume that the system works near the red sideband ($\Delta
=\omega _{m}$). In Fig. 2 we plot $T_{a}$ and $T_{c}$ as a
function of $\delta _{as}/\omega _{m}$ and $\delta _{cs}/\omega _{m}$,
respectively. Here we hold the pump power $P_{c}$ constant, $P_{c}$ $%
=100 $nW \cite{30}. We can see that the transmission of the left-moving
signal field is simply that of a bare resonator, and $T_{c}$ exhibits a dip
near $\delta _{cs}=\omega _{m}$ ($T_{c}\approx 0$). By adjusting the pump
power $P_{a}$, we find that the transmission of the right-moving signal
field can be obviously modified. Near $\delta _{as}=\omega _{m}$, $T_{a}$
will exhibit a very narrow peak and gradually increase with the increase of $%
P_{a}$. When $P_{a}=10^{4}P_{c}$, we have $T_{a}\approx 1$. If we continue
to increase $P_{a}$, the spectrum will exhibit a split, and this is
associated with the normal mode splitting \cite{31}. The above features can
result in a nonreciprocal transmission of the signal fields in our system,
i.e., the right-moving signal field is completely transmitted, while the
left-moving signal field is blocking-up.

Then we consider that the system works near the blue sideband ($\Delta
=-\omega _{m}$), and we also choose $P_{c}=100$nW. In Fig. 3, we can see
that $T_{c}$ exhibits a dip near $\delta _{cs}$ $=$ $-\omega _{m}$ ($%
T_{c}\approx 0$). By adjusting the pump power $P_{a}$, $T_{a}$ will exhibit
a very narrow peak near $\delta _{as}=-\omega _{m}$, with the increase of $%
P_{a}$, the peak value will first increase and then decrease. When $%
P_{a}=9.5P_{c}$, we have $T_{a}>1$, $T_{c}<1$, now the right-moving signal
field can be amplified, while the left-moving signal field cannot be
amplified. When $P_{a}=10^{4}P_{c}$, we have $T_{a}\approx 1$, $T_{c}\approx
0$. Now the system can also be used to realize the nonreciprocal
transmission of the signal fields.

In addition, in our system, the transmissive direction of the signal field
can be changed by adjusting the ratio of $P_{a}$ and $P_{c}$. For example,
when $\Delta =\omega _{m}$, $P_{a}=100$nW and $P_{c}=10^{4}P_{a}$, now the
left-moving signal field is completely transmitted while the right-moving
signal field is blocking-up.

\section{Nonreciprocal fast-slow light effects}

In this section, we show how to realize the nonreciprocal fast-slow light
propagation of the signal fields, i.e., both the right-moving and left-moving
signal fields can be completely transmitted, while the group velocity
of the right-moving signal field will be speed up and the left-moving signal
field will be slowed down, or vice versa.

In Fig. 4, we plot $T_{a}$ and $T_{c}$ for different intrinsic photon loss
rate $\kappa _{in}$ under the unbalanced-pumping condition ($%
P_{a}=10^{5}P_{c}$, $P_{c}=100$nW). We find that with the decrease of $%
\kappa _{in}$ the transmission of the left-moving signal field $T_{c}$ will gradually increase near $\delta
_{cs}=-\omega _{m}$. However, the transmission of the right-moving signal field $T_{a}\approx 1$ near $\delta _{as}$ $=$ $-\omega _{m}$. When $\kappa _{in}=10^{-1}\kappa $, we have $%
T_{a}\approx 1$, $T_{c}\approx 0.7$, now the nonreciprocity of the system
about the optical transmission is weakened. When $\kappa _{in}=10^{-3}\kappa 
$, we have $T_{a}\approx 1.01$, $T_{c}\approx 0.996$, and the nonreciprocity of the system about 
the optical transmission almost disappears（($T_{a}\approx T_{c}$).

However, in this situation ($\kappa _{in}\ll \kappa $), the nonreciprocity
of system is shown in the group delay properties of the signal fields.
In Fig. 5, we plot $\tau _{a}$ and $\tau _{c}$ as a function of $\delta
_{as}/\omega _{m}$ and $\delta _{cs}/\omega _{m}$, respectively. We can see
that in the range of the parameters we considered (we have plotted the
transmission spectra $T_{a}$ and $T_{c}$ and we can guarantee that $%
T_{a}\approx T_{c}\approx 1$ for all the parameters used in Fig. 5), the
group delay of the right-moving signal field is negative near $\delta
_{as}=-\omega _{m}$ (the group velocity will be speed up), that corresponds
to the fast light propagation. While the group delay of the left-moving
signal field is positive near $\delta _{cs}=-\omega _{m}$ (the group
velocity will be slowed down), that corresponds to the slow light
propagation. This shows that the system can  exhibit a nonreciprocal fast-slow light
propagation of the signal fields. 

Furthermore, we can change the propagation direction of the
fast-slow light by adjusting the ratio of of $P_{a}$ and $P_{c}$. For
example, in Fig. 5(c), we have $\tau _{a}\approx -0.3\mu s$ and $\tau
_{c}\approx 0.3\mu s$. However, if we choose $P_{a}=100$nW and $%
P_{c}=10^{5}P_{a}$, then we have $\tau _{a}\approx 0.3\mu s$ and $\tau
_{c}\approx -0.3\mu s$, now the right-moving signal field is slow light and
the left-moving signal field is fast light.

\section{Conclusion}

In summary, we have studied the nonreciprocal transmission and the fast-slow
light effects in a cavity optomechanical system. We have shown that for both
the red-sideband pumping or the blue-sideband pumping, the system can act
as an optical unidirectional isolator. We have also shown that if the
intrinsic photon loss is much less than the external coupling loss, the
nonreciprocity of the system on the optical transmission almost disappears,
now the system reveals an interesting nonreciprocal fast-slow light
propagation phenomenon. Our proposed model might have applications in the photonic network.

\textbf{Fundings.} This work was supported by the National Natural Science
Foundation of China (Nos. 11574092, 61775062, 61378012, 91121023); the
National Basic Research Program of China (No. 2013CB921804).

\section*{Appendix}

By substituting Eq. (5) into Eqs. (4), we can obtain the the steady-state
mean value equations 
\begin{align}
0=& -(i\Delta +\kappa _{t})A_{0}-igA_{0}(B_{0}+B_{0}^{\ast
})-iJC_{0}+\varepsilon _{ap},  \notag \\
0=& -(i\Delta +\kappa _{t})C_{0}-igC_{0}(B_{0}+B_{0}^{\ast
})-iJA_{0}+\varepsilon _{cp},  \notag \\
0=& -(i\omega _{m}+\gamma )B_{0}-ig(\left\vert A_{0}\right\vert
^{2}+\left\vert C_{0}\right\vert ^{2}).  \tag{A1}
\end{align}%
In this system, we are interested on the dynamics of the cavity mode
components $A_{a+}e^{-i\delta _{as}t}$ and $C_{c+}e^{-i\delta _{cs}t}$ which
are resonance with the corresponding signal fields $\varepsilon
_{as}e^{-i\delta _{as}t}$ and $\varepsilon _{cs}e^{-i\delta _{cs}t}$,
respectively. We can obtain 
\begin{align}
\Phi _{a}B_{a+}=& -ig(A_{0}^{\ast }A_{a+}+A_{a+}^{\ast }A_{0}+C_{0}^{\ast
}C_{a+}+C_{a+}^{\ast }C_{0}),  \notag \\
\Omega _{a}A_{a+}=& -igA_{0}(B_{a+}+B_{a+}^{\ast })-iJC_{a+}+\varepsilon
_{as},  \notag \\
\Omega _{a}C_{a+}=& -igC_{0}(B_{a+}+B_{a+}^{\ast })-iJA_{a+},  \tag{A2} \\
\Phi _{c}B_{c+}=& -ig(A_{0}^{\ast }A_{c+}+A_{c+}^{\ast }A_{0}+C_{0}^{\ast
}C_{c+}+C_{c+}^{\ast }C_{0}),  \notag \\
\Omega _{c}C_{c+}=& -igC_{0}(B_{c+}+B_{c+}^{\ast })-iJA_{c+}+\varepsilon
_{cs},  \notag \\
\Omega _{c}A_{c+}=& -igA_{0}(B_{c+}+B_{c+}^{\ast })-iJC_{c+},  \tag{A3}
\end{align}%
where $\Phi _{k}=i(\omega _{m}-\delta _{ks})+\gamma $ and $\Omega
_{k}=i[\Delta +g(B_{0}+B_{0}^{\ast })-\delta _{ks}]+\kappa _{t}$ , $k=a,c$.


\begin{thebibliography}{99}
\bibitem{30} It should be point out that $P_{c}$ cannot be zero, since the
	linearization demands a strong the pump fields. For the parameters here, we
	take $P_{c}=$ $100$ nW, and we have $C_{0}\approx 28$.
\end{thebibliography}
\end{document}